\documentclass[aps,twocolumn,showpacs]{revtex4}
\usepackage{dcolumn}
\usepackage{graphicx}
\usepackage{amsmath}
\usepackage{amsfonts}
\usepackage{amssymb}
\usepackage{psfrag}
\usepackage{wrapfig}
\usepackage{subfigure}
\usepackage{makeidx}
\usepackage{bm}
\usepackage{epsf}
\DeclareMathOperator{\sech}{sech}

\begin{document}

\title{Atomic Bright Soliton Interferometry}
\author{Li-Chen Zhao$^{1,2}$}\email{zhaolichen3@nwu.edu.cn}
\author{ Guo-Guo Xin$^{1,2}$}\email{xingg@nwu.edu.cn}
\author{ Zhan-Ying Yang$^{1,2}$}
\author{Wen-Li Yang$^{1,2,3}$}
\address{$^{1}$School of Physics, Northwest University, Xi'an, 710069, China}
\address{$^{2}$Shaanxi Key Laboratory for Theoretical Physics Frontiers, Xi'an, 710069, China}
\address{$^{3}$Institute of Modern Physics, Northwest University, Xi¡¯an 710069, China}

\date{\today}
\begin{abstract}

The properties of nonlinear interference pattern between atomic bright solitons are characterized analytically, with the aid of exact solutions of dynamical equation in mean-field approximation. It is shown that relative velocity, relative phase, and nonlinear interaction strength can be measured from the interference pattern. The nonlinear interference properties are proposed to design atomic soliton interferometry in Bose-Einstein condensate. As an example, we apply them to measure gravity acceleration in a ultra-cold atom systems with a high precision degree.  The results are also meaningful for precise measurements in  optical fiber, water wave tank, plasma, and other  nonlinear systems.

\end{abstract}
\maketitle

\section{Introduction}
Interference is a fundamental property of both classical and quantum wave objects. The simplest case is the linear interference of two plane waves, which possess $k_1$ and $k_2$ wave vectors separately. The spatial interference period is obviously $\frac{2 \pi}{|k_1-k_2|}$. This character holds for all linear interference which comes from linear superposition principle. But it is unusual to investigate interference of plane wave in real physical systems, since there is no precise plane wave in the systems. Usually, the interference process is observed based on wave packets \cite{Kovachy}. A wave packet is a superposition of plane waves possessing many different wave vectors, and it will disperse with time evolution. The multi-wave vector differences induce many different spatial interference periods, which makes the linear interference pattern admit low visibility. These characters would affect the precision of matter wave interferometry. Notably,   soliton in Bose-Einstein
condensate (BEC)  is a special wave packet which does not disperse \cite{Zabusky,Nguyen,Marchant,Medley}, which could admit much higher visibility fringes than the linear wave packets \cite{Billamw}. Recently, bright matter wave solitons' interference fringe was demonstrated experimentally \cite{McDonald}. The experiment indeed suggested that solitonic matter wave significantly increased fringe visibility  compared with non-interacting atomic clouds.

Soliton-based matter-wave interferometer was also proposed theoretically in a harmonic potential trap with a Rosen-Morse barrier at its
center \cite{Polo}.  Sagnac interferometry using bright matter-wave solitons was  proposed based on the bright solitons colliding with a barrier on a ring \cite{Helm}. Moreover, as shown in the experiment proposed in \cite{Boixo,Boixo1},  a two-mode BEC of $N$ atoms is used to implement a nonlinear Ramsey interferometer whose detection
uncertainty scales better than the optimal $1/N$ Heisenberg scaling of linear interferometry. Recent studies further suggested that  the pairwise scattering interaction in a BEC would make nonlinear atom interferometer surpass classical precision limit greatly \cite{Tacla,Gross}. For example, the interference of BEC was used to probe submicron forces \cite{Dimopoulos}.  However, the interference process with pairwise scattering interaction is a nonlinear one, for which the explicit laws are hard to described as clearly as linear interference. Therefore, it is essential to characterize  the definite laws well for the nonlinear interference pattern, which can be used to measure some physical parameters more conveniently and precisely. We have partly characterized the properties of periodic behaviors in the interference processes \cite{zhaofer}. We would like to further clarify the interference properties (the relative phase's role is addressed clearly), and demonstrate applications of them in details to design atomic soliton interferometry.

In this paper, we  characterize the properties of interference pattern between two atomic bright solitons in details,  and compare them with the linear interference properties. The spatial period does not depend on nonlinear strength and it is determined by the relative effective wave vector between solitons,  if the soliton states are admitted by the nonlinear systems.  This character is identical with spatial period properties of linear interference. Moreover,  the temporal period of interference pattern is characterized clearly, which is different from the ones in linear interference cases. And this can be used to read out nonlinear coefficient from the interference pattern.  Furthermore, we demonstrate that relative phase between solitons can be read out from the interference pattern. These results can be used to design atomic soliton interferometry.  As an example, we discuss how to  measure gravity acceleration based on matter wave solitons.

\section{Theoretical results of atomic soliton interferometry}
In recent years, BECs have been proven to be appealing systems for the realization of
atom interferometers, owing to their properties of macroscopic phase coherence \cite{Simon,Muntinga,Dickerson,Ahlers,Cronin}. Recently, a method was proposed  to split the ground state of an attractively interacting atomic Bose-Einstein condensate into two bright solitary waves with controlled relative phase and velocity \cite{Billam}. Soliton interaction in BEC systems have been demonstrated widely in real experiments \cite{Nguyen,Marchant,Medley}. Those experiments inspire us to propose an atomic bright soliton interferometry based on the interference properties. Then, we firstly investigate the general properties of interference pattern which emerges during bright solitons overlapping.

 For a cigar-shaped BEC in a harmonic trap, the state of BEC can be described by a normalized condensate mode-function in mean-field approximation, and its dynamical equation is
$ i \hbar\frac{\partial U(x,t)}{\partial t}+\frac{\hbar^2}{2m}\frac{\partial^2
U(x,t)}{\partial x^2}+2 \gamma_{1D} |U(x,t)|^2 U(x,t)+\frac{\omega^2 x^2} {2}  U(x,t) =0$, where $\gamma_{1D}= 2 h \omega_r a$ ( $a$ is the s-wave scattering length ), $\omega$ is the axial frequency in our units of inverse time. When $\omega << 1$, the harmonic trap potential will have little effects on soliton dynamics and can be neglected. With dimensionless unit, the dynamical equation can be written as follows,
\begin{equation}
i\frac{\partial U(x,t)}{\partial t}+\frac{\partial^2
U(x,t)}{\partial x^2}+2 \gamma |U(x,t)|^2U(x,t) =0,
\end{equation}
which has been studied widely in soliton fields. If $\gamma=0$, the equation will become a linear Sch\"{o}dinger equation, which can be used to discuss the linear interference directly. The simplest nontrivial solution is $U(x,t)= a e^{ikx-i k^2 t}$, the interference of two plane wave will be $a_1  e^{ik_1\ x-i k_1^2 t+i \phi}+ a_2  e^{ik_2\ x-i k_2^2 t}$. Obviously, the periodic properties of interference term are reflected by  $\cos[(k_1-k_2)x+(k_1^2-k_2^2)t+\phi]$. One can see that the spatial period is $\frac{2 \pi} {|k_1-k_2|}$. We will demonstrate that this character holds for all bright soliton interference process, and does not depend on the nonlinear strength. The temporal period is $\frac{2 \pi} {|k_1^2-k_2^2|}$. This character will be revised for a nonlinear interference process. Moreover, the relative phase is hard to be read out from the linear interference pattern, since there is an infinite periodic fringe, the center is hard to be defined and located. However, this defect will be solved perfectly by the following nonlinear interference process.

If $\gamma>0$, the BEC system will admit bright soliton which is a hump density on zero background; if $\gamma<0$, the BEC system will admit dark soliton which is a defect on plane wave background. We firstly discuss the interference of bright solitons. For simplicity without losing generalities, we discuss the interference properties of two bright solitons. The initial two well-separated  solitons (distance between soliton $|x_{10}-x_{20}|$ is much larger than the soliton scale) can be written as $\sqrt{p_1} \ \sech[\sqrt{\gamma p_1} (x-x_{10}-v_1 t)] \ e^{iv_1 (x-x_{10})/2-iv_1^2 t/4+i \gamma p_1 t} \ e^{i \phi_1}+ \sqrt{p_2} \ \sech[\sqrt{\gamma p_2} (x-x_{20}-v_2 t)] \ e^{iv_2 (x-x_{20})/2-iv_2^2 t/4+i \gamma p_2 t} e^{i \phi_2}$, for which the parameters $v_j$ ($j=1,2$) denote the velocities of solitons respectively, $p_j$ denote solitons' density peak values, and $\phi_1 -\phi_2 = \phi \in [0, 2 \pi]$ denotes the relative phase between them. Especially, each soliton has an effective wave vector (EWV), which is unique and can be used to characterize the interference property conveniently. The EWV is defined as mean value of  wave vectors for atoms in a soliton. Therefore, the EWV of a soliton is $k_j=v_j/2$.

\begin{figure}[htb]
\centering
\label{fig:1}
{\includegraphics[height=72mm,width=85mm]{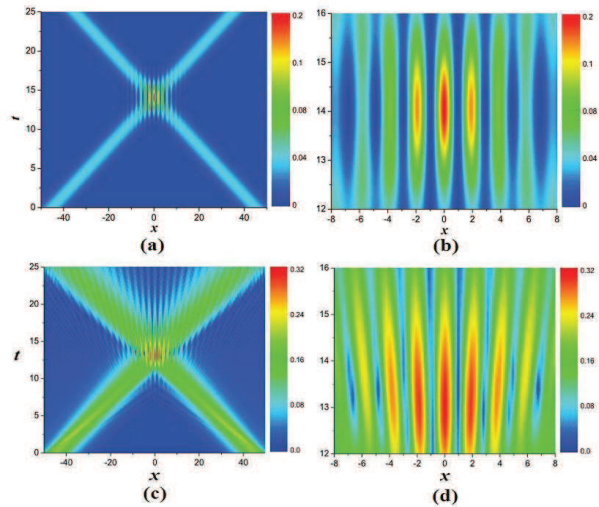}}
\caption{(color online) (a) and (b): The density plot of nonlinear interference process between
the two atomic solitons. It is shown that the spatial interference pattern
appears in the collision region when the relative velocity between
solitons is large. The parameters are $\gamma=1, p_1=p_2=0.04, v_1=3.2, v_2=-3.2 $, and $\phi=0$. (c) and (d): The linear interference process between wave packets, evolved from two initial soliton states. The numerical evolution is given by the dynamical equation with $\gamma=0$ from the  two initial soliton states with identical parameters with (a) and (b). It is seen that the linear interference pattern become irregular and the pattern visibility becomes lower than the nonlinear interference pattern.}
\end{figure}

The linear superposition form just holds well for the large separation between solitons. If the solitons approach each other and overlap in a certain extent, the nonlinear interaction will make the linear superposition form fail to describe the evolution dynamics of solitons. But the interaction between them can be described by the well-known two-soliton solution \cite{zhaofer}. The interaction between more than two solitons can be studied based on multi-soliton solution which can be obtained by B\'{a}cklund transformation directly \cite{Mat,Cies,Dok}. Based on the exact two-soliton solution, we can derive that the
property of spatial interference pattern, namely
\begin{equation}
D=\frac{2\pi}{|k_1-k_2|},
\end{equation}
where $k_j=v_j/2$ is the EWV of soliton. This character is similar to the interference of plane waves in linear case. Spatial
interference pattern can be observed when the relative EWV is large enough that
soliton's matter wavelength is smaller than scale of
solitons \cite{Snyder,Kumar}, such as Fig.\ 1 (a). When the matter wave length is not smaller than the scale of soliton, namely, solitons' relative EWV is small, one could not observe the interference pattern.

We emphasize that the spatial period does not depend on the nonlinear strength for solitons admitted by the corresponding nonlinear system. It has been shown that a localized wave packet evolves to be a soliton or multi-soliton admitted by system \cite{SFWP}. We consider the initial condition $\sqrt{p_1} \ \sech[\sqrt{\gamma p_1} (x-x_{10}-v_1 t)] \ e^{iv_1 (x-x_{10})/2-iv_1^2 t/4+i \gamma p_1 t} \ e^{i \phi_1}+ \sqrt{p_2} \ \sech[\sqrt{\gamma p_2} (x-x_{20}-v_2 t)] \ e^{iv_2 (x-x_{20})/2-iv_2^2 t/4+i \gamma p_2 t} e^{i \phi_2}$ with $\gamma=1$,  and evolve it numerically by the Eq. (1) with different $\gamma$ values.  We can see that the interference patterns  with $\gamma \neq 1$ changes  comparing with the case with $\gamma=1$,  because the packets have not evolved to be a stable soliton state when they collide each other. However, if the solitons collide with each other after the wave packets have evolved to be stable solitons admitted by the system, the spatial period of interference pattern is still independent of other factors but relative velocity. Especially, if the nonlinear coefficient $\gamma=0 $, we evolve the initial condition by Eq. (1). The soliton packet will disperse with time evolution, and the visibility of interference pattern will be reduced greatly.  The results are shown in Fig. 1 (c) and (d) with the identical initial condition of Fig. 1 (a) and (b). The results show that soliton interferometry just apply to nonlinear cases, and it has an obvious merit in contrast to interference between wave packets in linear cases. Similar expansion interference has been demonstrated experimentally in a BEC with repulsive interactions between atoms \cite{Andrews,Schumm}.  It is noted that the nonlinear interference spatial period has identical form with the interference of two plane waves in linear cases. Besides this character, soliton has quite different properties from linear localized waves. Linear localized wave packets usually disperse and they interference pattern usually admit low visibility. The stability of soliton can be used to improve the visibility and sensitivity of interferometry greatly \cite{McDonald,Billamw}.

Moreover, there is an oscillating behavior along time evolution, if the peak values are not equal in the case in Fig. 1, or the absolute values of velocities are not equal.  Based on generalized two-soliton solutions, we derive an exact expression for temporal interference period
\begin{equation}
T=\frac{2\pi}{\gamma (p_2-p_1)+k_1^2-k_2^2}.
\end{equation}
Namely, the temporal-period is determined by both the peaks and EWVs
of solitons, and the nonlinear coefficient. These characters are different from the temporal period for linear interference case. Therefore, the temporal period can be used to measure nonlinear coefficient in many different physical systems. If one soliton's velocity is zero,
the  parameter $\gamma$ can be given as
\begin{equation}
\gamma=\frac{2\pi}{ T (p_2-p_1)}+\frac{4 \pi^2}{D^2(p_2-p_1)}.
\end{equation}
The nonlinear parameter is related with scattering length directly for ultra-cold atoms. This can be used to measure the scattering length.  Especially,  the temporal interference period will be $T=\frac{2\pi}{|k_1^2-k_2^2|} $ for solitons with $p_1=p_2$. We can see that the nonlinear interference properties of solitons with identical density peaks agree perfectly with the ones of linear interference process. In this case, the nonlinear parameter can not be read out from the interference pattern anymore.

\begin{figure}[htb]
\centering
\label{fig:3}
{\includegraphics[height=50mm,width=75mm]{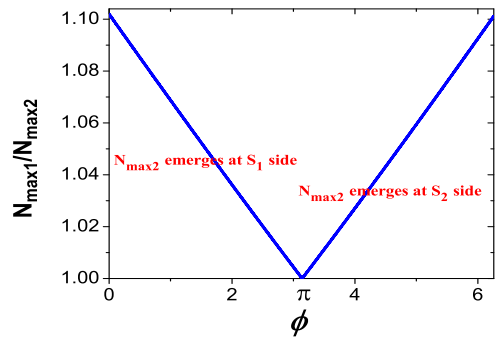}}
\caption{(color online) The relation between ratio of main maximum density value and secondary maximum value and relative phase $\phi$. The parameters are $\gamma=1, p_1=p_2=0.04, v_1=4, v_2=-4 $. }
\end{figure}

The interference pattern's periodic properties is not related with the relative phase between atoms. It should be pointed that we  addressed the above nonlinear interference properties before \cite{zhaofer}. But the relative phase affects the density distribution profile and has not been discussed in details. There is only one maximum density value and the pattern is symmetric on spatial distribution direction for the relative phase is zero, but there are two equal maximum density values for the relative phase is $\pi$. There interference pattern will become asymmetry for other relative phase values.  A recent experiment also showed that collisions of matter-wave bright solitons depended crucially on their relative phase \cite{Nguyen}. This can be used to test relative phase between solitons. Furthermore, we find a way to read out the relative phase $\phi$ between solitons.
The ratio of main maximum density value (denoted by $N_{max1}$) and secondary maximum value (denoted by $N_{max2}$) can be used to reflect the relative phase precisely. With the $p_1=p_2=p$ and $v_1=-v_2=v$, the relation can be derived for $\phi \in [0,\pi]$ in a simple analytical form as
\begin{equation}
\frac{N_{max1}}{N_{max2}}=\ \cosh[\frac{a (\phi-2 \pi)} { k}]^2\ \sech[\frac{a \phi} { k}]^2,
\end{equation}
where $a=\frac{\sqrt{\gamma p}}{2}$, $k=v/2$. The relation is $\frac{N_{max1}}{N_{max2}}=\ \cosh[\frac{a \phi} {k}]^2\ \sech[\frac{a (\phi-2 \pi)} {k}]^2 $ for $\phi \in [\pi,2 \pi]$.  The explicit relations between them can be described by the line in Fig. 2. This means that we just measure the density value of $N_{max1} $ and $N_{max2}$, and see which side the secondary maximum density value locates. Then the relative phase can be calculated.  It should noted that  the relative phase is defined as $ \phi=\phi_1 -\phi_2  \in [0, 2 \pi]$ ($\phi_j$ denotes soliton $j$ 's phase). If the  $N_{max2}$ locates on the soliton $1$ side, the relative phase can be known the left  line segment in Fig. 2; if the  $N_{max2}$ locates on the soliton $2$ side, the relative phase can be known the right line segment in Fig. 2. When the $N_{max1}=N_{max2}$, we can know the relative phase is $\pi$ from the line in Fig. 2. This way for measuring relative phase between solitons is different from the bright soliton interferometers based on the phase-dependent soliton recombination at a potential barrier \cite{Polo,Helm,Boris}.

The general properties of interference pattern are clarified analytically based on the well-known exact two-soliton solution of nonlinear Schr\"{o}dinger equation. It is shown that the relative phase, relative velocity, and nonlinear interaction strength can be measured from the interference pattern between solitons.  This enables us to propose atomic soliton interferometry, which can be used measure physical quantities more precisely and conveniently for its higher visibility and lower thermal noises. After splitting the ground state of an attractively interacting atomic Bose-Einstein condensate into two bright solitary waves with controlled relative phase and velocity \cite{Billam}, we can design proper operations on each soliton by the quantity to be measured  as done in \cite{Biedermann,Muller,Dimopoulos2,Asenbaum}. Then we can measure the interference pattern by recombing the two solitons.  In this way, the quantity will be calculated by the above analytical results. The interference pattern can be  used to hold great promise
for precision measurements \cite{Polo,22}, including measurements
of gravity \cite{23,24,25,26}, rotations and magnetic field
gradients \cite{29}, and other quantum superpositions \cite{30,31}.  For an example, we show how to measure gravity acceleration based on a two-soliton interference pattern.

\begin{figure}[htb]
\centering {\includegraphics[height=85mm,width=85mm]{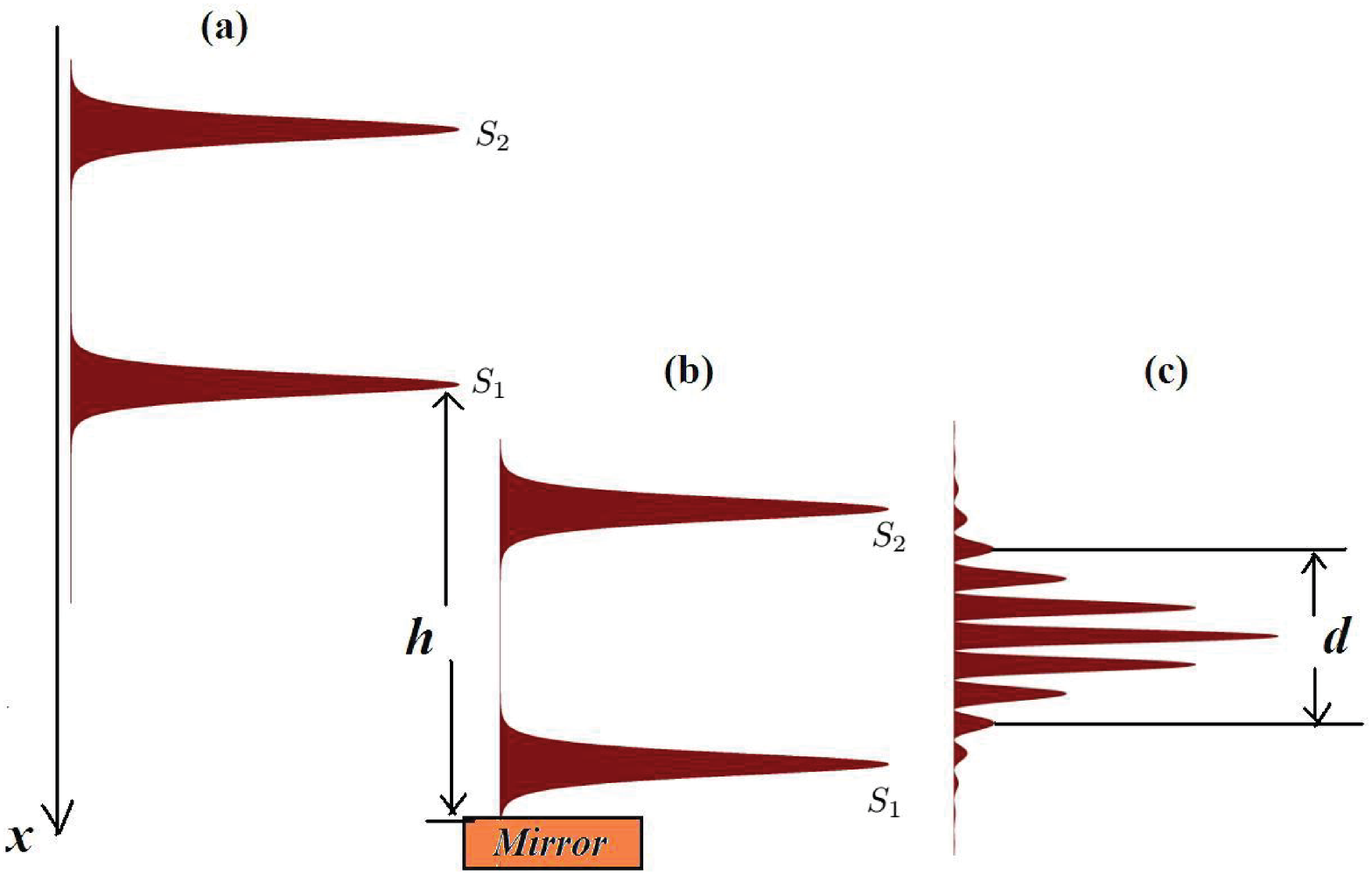}}
\caption{(color online) A simple schematic diagram for measuring gravity acceleration based on soliton interference pattern. (a) Locate two solitons with zero initial velocities and identical profiles along gravity acceleration direction.  (b) The first soliton $S_1$ arrive at the mirror and be reflected back. (c) The two solitons approach and overlap each other. The interference pattern can be imaged by optical method. Measure the distance between the mirror and the initial location of $S_1$ soliton, and measure $n$ times spatial period distance as $d$. The gravity acceleration will be calculated directly from the interference properties. }
\end{figure}

\section{Application to measure gravity acceleration}
For a cigar-shaped BEC in a harmonic trap with a gravity field along $x$ direction, the state of BEC can be described by a normalized condensate mode-function in mean-field approximation, and its dynamical equation is
$ i \hbar\frac{\partial U(x,t)}{\partial t}+\frac{\hbar^2}{2m}\frac{\partial^2
U(x,t)}{\partial x^2}+2 \gamma_{1D} |U(x,t)|^2 U(x,t)+(\frac{\omega^2 x^2} {2} + m g x )U(x,t) =0$, where $\gamma_{1D}= 2 h \omega_r a$ ( $a$ is the s-wave scattering length ), $\omega$ is the axial frequency in our units
of inverse time. When $\omega << 1$, the harmonic trap potential will have little effects on soliton dynamics and can be neglected. Interestingly, the dynamical equation with linear potential is still integrable and admits exact soliton solution \cite{cplzhao}. The solution can be derived exactly and the explicit expressions are not shown here. Based on the soliton solution, we can see that the soliton profile is identical with the ones of standard nonlinear Schr\"{o}dinger equation Eq. (1) and it has an acceleration $g$ along $x$ direction. Then we design a way to measure the gravity acceleration based on the above interference properties.We locate two solitons (denoted by $S_1$ and $S_2$) initially with zero velocity and identical profile on arbitrary two sites  along $x$ direction. Set a mirror at the site which is $h$ far from the first soliton location. The mirror is refer to some proper settings for reflecting the soliton \cite{Marchant}. For an example, an electron beam can be used to reflect the soliton very well \cite{Xzhang}. The two solitons will move along $x$ under the influence of gravity field. The first soliton will be reflected on the mirror. Then the two solitons will approaching each other and the interference pattern can be imaged by optical testing method. One can easily prove that the relative velocity between them is $2\ \sqrt{2 g h} $ and be unvaried for the whole interference process.  Then measure the distance between two interference peaks (denoted by $d$), and count the peak number between them (denoted by $n-1$). The distance $d$ includes $n$ spatial periods, namely, $d=n\ D$.  A simple schematic diagram is shown in Fig. 3. Based on the general property for spatial interference,  we can calculate the gravity acceleration as
\begin{equation}
g=\frac{2 n^2 \pi^2}{d^2 h}.
\end{equation}
 Based on the experimental results on BEC life time (about $50\  sec$ as shown in \cite{Hamner})  and spatial measurement precision (about $0.1\  \mu m$ in \cite{Simsarian}), we can design the operation to choose $h=0.5 \ m \pm 10^{-9} m$ and $d=1\ mm \pm 10^{-7} m $. There are about $n~500$ interference stripes in the $1\ mm$ region, and the number $n$ can be increased to improve the precision of $D$. We can evaluate the precision of $g$ is about $ 10^{-10} g $. The precision can be further improved by longer BEC lifetime or more precise spatial measurement on interference stripe. The visibility of soliton interference pattern is much higher than the interference process of two wave packets observed in  \cite{Kovachy}, since dispersive effect plays important role in the previously reported interference process. Additionally, the interference property on relative phase can be used to check whether there is a $\pi$ phase shift for matter wave soliton reflecting on a mirror pulse, since the relative phase between soliton can be also read out from the interference pattern.

\section{Conclusion and discussion}
In summary, we fully characterize the properties of nonlinear interference pattern between atomic bright solitons. The EWV of soliton is defined reasonably to describe the spatial period and temporal period of interference pattern. It shown that relative velocity, relative phase, and nonlinear interaction strength can be measured from the interference pattern. The nonlinear interference properties are proposed to design atomic soliton interferometry, which is different from the bright soliton interferometers based on the phase-dependent soliton recombination at a potential barrier \cite{Polo,Helm,Boris}. As an example, we apply them to measure gravity acceleration in a ultra-cold atom systems, which demonstrate a high precision degree.  The atomic soliton interferometry will be available  in the near future by using current technologies on matter wave solitons in BEC systems \cite{Billam,Nguyen,Marchant,Medley}. Especially, with a very small repulsive three-body interaction, stable three dimensional quantum balls can exist and interfere with each other, as shown in \cite{3DS}. The interference properties could be also used to design nonlinear interferometry.

It is well known that bright soliton  can not exist with repulsive case and no trapping potentials. BEC with repulsive interaction usually admits dark soliton which is a defect on plane wave background.  Detail investigation indicates that there is no interference pattern for scalar dark solitons. However, for vector solitons in multi-component BEC \cite{BBD}, we expect that dark solitons in one component can admit interference pattern, which is associated with bright solitons' interference behavior in other components. The results can be extended to study interference pattern of other nonlinear localized waves, such as vector bright soliton, bright-dark soliton, etc. Moreover, the soliton interferometry could be also extended to other physical systems, such as optical fiber, water wave tank, and even plasma systems \cite{Onorato}.

\section*{Acknowledgments}
 This work is supported by National Natural Science Foundation of
China (Contact No. 11775176), Shaanxi Association for Science and Technology (Contact No. 20160216).

\end{document}